\documentclass[journal,final]{IEEEtran}
\usepackage{graphicx}

\begin{document}

\title{Analytical determination of participation in superconducting coplanar architectures}

\author{Conal E. Murray,
        Jay M. Gambetta,
	Douglas T. McClure and 
	Matthias Steffen
\thanks{C.E. Murray, J.M. Gambetta, D.T. McClure and M. Steffen are with International Business Machnies,
IBM T.J. Watson Research Center, Yorktown Heights,
NY, 10598 USA  (e-mail: conal@us.ibm.com).}
}

\maketitle

\begin{abstract}
Superconducting qubits are sensitive to a variety of loss mechanisms which include dielectric loss from interfaces. The calculation of participation near the key interfaces of planar designs can be accomplished through an analytical description of the electric field density based on conformal mapping. In this way, a two-dimensional approximation to coplanar waveguide and capacitor designs produces values of the participation as a function of depth from the top metallization layer as well as the volume participation within a given thickness from this surface by reducing the problem to a surface integration over the region of interest.  These quantities are compared to finite element method numerical solutions, which validate the values at large distances from the coplanar metallization but diverge near the edges of the metallization features due to the singular nature of the electric fields.  A simple approximation to the electric field energy at shallow depths (relative to the waveguide width) is also presented that closely replicates the numerical results based on conformal mapping and those reported in prior literature.  These techniques are applied to the calculation of surface participation within a transmon qubit design, where the effects due to shunting capacitors can be easily integrated with those associated with metallization comprising the local environment of the qubit junction.
\end{abstract}
\begin{IEEEkeywords}
Conformal mapping, electromagnetic simulation, planar structures, quantum devices, coplanar waveguides
\end{IEEEkeywords}

\IEEEpeerreviewmaketitle

\section{Introduction}

\IEEEPARstart{D}{ecoherence} in superconducting qubits can be caused by dielectric loss generated on many surfaces in and around the environment of the qubit.  For example, contamination layers and native oxides present on semiconductor and metallization surfaces will exhibit participation based on the amount of energy induced by the electric fields produced by the qubit system.  Traditionally, finite element method (FEM) based models of the transmon qubit architectures are employed to estimate surface participation from interfaces between the substrate and metallization (SM), the free surface of the substrate (SA) and the top surface of the metallization (MA) \cite{Wenner11}, \cite{Sandberg12}, \cite{Wang15}, \cite{Calusine18}.  However, the large difference between the thicknesses of such layers and the length scales associated with the overall qubit or resonator design makes accurate calculation of surface participation in these domains difficult.  One method to circumvent this issue involves calculating the electric field density on the surfaces of interest and determining surface participation by modifying the appropriate components of the electric field based on the difference in dielectric constants between the interfacial layer and substrate \cite{Wenner11} - \cite{Gambetta16}.  The values obtained with this approach can be highly dependent on the discretization scheme used to tessellate the various structures due to the singular behavior of the electric fields near the corners and edges of conductors.  Hybrid schemes that can involve power law approximations have also been proposed \cite{Wenner11}, \cite{Wang15} to account for divergences in the electric field distributions.

A more robust method to calculate surface participation can be implemented based on a two-dimensional, analytical formulation of the electric field distributions generated by metallization features on dielectric substrates \cite{Wen69}, \cite{Gillick93}.  This approach, which assumes that the metallization is composed of two sheets and is perfectly conducting, incorporates the singular nature of the electric fields \cite{Jackson75} which scale as $r^{-0.5}$ where r is the distance from the metallization edges.  It will be shown that a closed form solution of the electric field energy can be generated that does not diverge and is applicable to both resonator and capacitor designs.

\begin{figure}[htbp!]
		 \centering
	\includegraphics[width=0.35\textwidth]{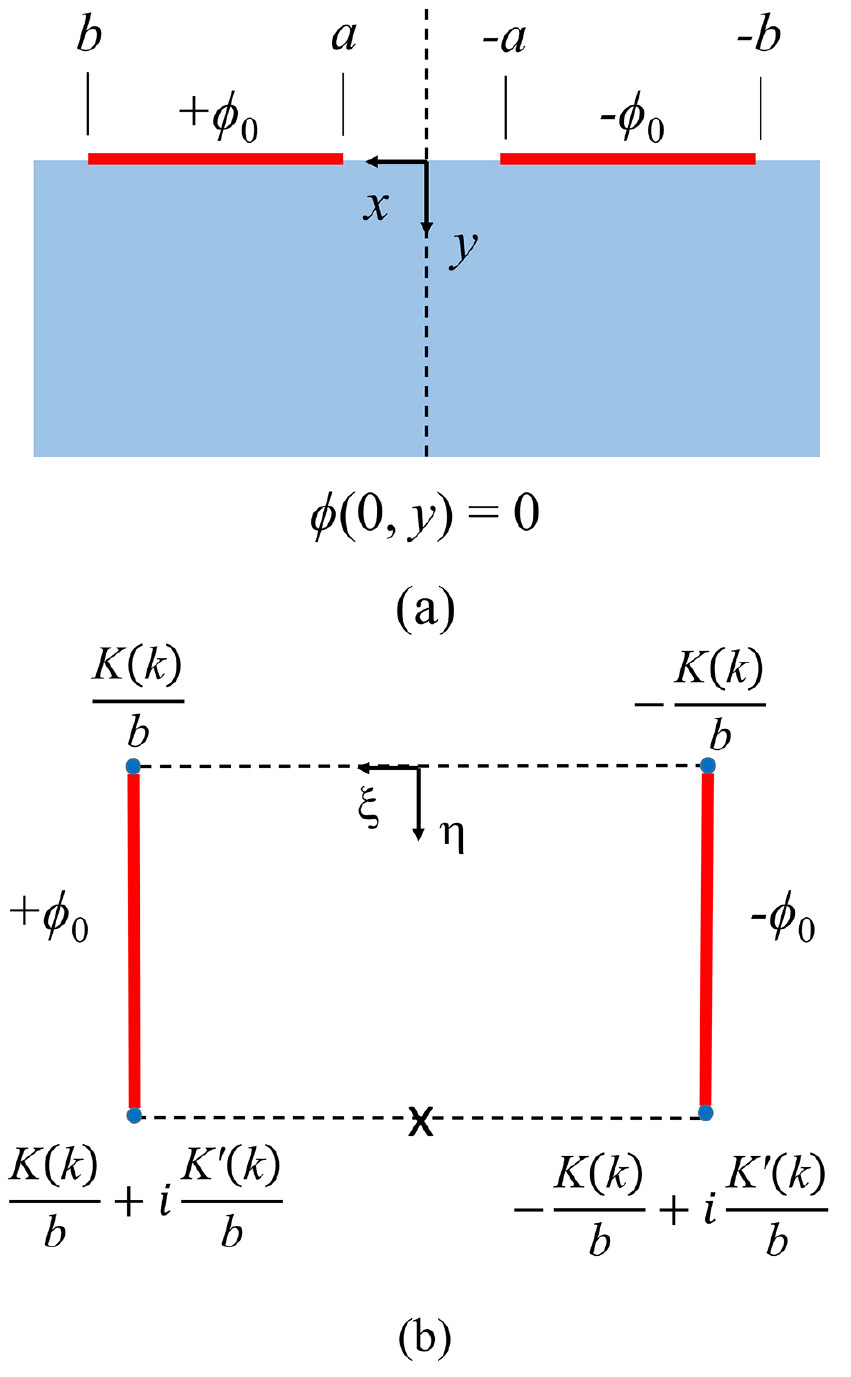}
		 \caption{(a) Cross-sectional schematic of a two-dimensional paddle metallization deposited on a semi-infinite substrate. (b) Transformed geometry through conformal mapping from the $x-y$ plane to the $\xi-\eta$ plane using (\ref{eq:eqcon1}).  The x on the bottom dotted line denotes the point corresponding to $-\infty$ and $+\infty$ along the $x$-axis from Fig. \ref{fig:padconform}a.}
		 \label{fig:padconform}
\end{figure}

\section{Coplanar capacitors}  \label{section:paddles}

Let us first analyze the shunting capacitors associated with a transmon qubit \cite{Gambetta16}, \cite{Koch07}, \cite{Rigetti12}.
A cross-sectional schematic of the geometry is shown in Fig. \ref{fig:padconform}, where we assume a semi-infinite substrate upon which two, perfect electrical conductor (PEC), metallization features with zero thickness reside.  This approach ignores the effects of finite metallization thickness and substrate recess due to etching, which can impact overall capacitance \cite{Yang98} and loss \cite{Bruno15} in resonators.  Under a quasi-static approximation, an electrostatic condition, in which opposite potentials $+\phi_0$ and -$\phi_0$ exist on the features, can be used to calculate the electric field energy.  This scenario is analogous to the odd mode generated along a coplanar stripline design.  Based on conformal mapping \cite{Wen69}, \cite{Gao08}, the structure in Fig. \ref{fig:padconform}a can be transformed from the complex $z$-plane
$(z = x + i y)$ to the w-plane $(w = \xi + i \eta)$ into a parallel-plate capacitor (Fig. \ref{fig:padconform}b):
\begin{equation} \label{eq:eqcon1}
w = \int{\frac{dz}{\sqrt{(z^2-a^2)(z^2-b^2)}}}
\end{equation}
where $a$ is half of the distance between the metallization features and $b$ is half of the distance between the outer edges of the metallization features.  On the $w$-plane, the metallization width is equal to $\frac{K'(k)}{b}$ with $K'(k) = K(k')$ referring to the complement of the complete ellitpic integral of the first kind, $k' = \sqrt{1-(\frac{a}{b})^2}$. The form of the electric field in Fig. \ref{fig:padconform}b is aligned parallel to the $\xi$ axis and is equal to the difference in potential ($2\phi_0$) divided by the distance between the metallic features $\frac{2 K(k)}{b}$ where $k = \frac{a}{b}$:
\begin{equation}\label{eq:efieldw}
E_\xi - i E_\eta = -\frac{\phi_0 b}{K(k)}
\end{equation}
which can also be written in the form of a complex potential:
\begin{equation}\label{eq:phiw}
\phi(\xi) = \frac{\phi_0 b}{K(k)} \xi
\end{equation}
To transform this electric field back into the $z$-plane, we can use the following relation \cite{Gillick93}
\begin{equation}\label{eq:phider}
E_x - i E_y = -\frac{\partial \phi}{\partial z} = -\frac{\partial \phi(\xi)}{\partial \xi} \frac{\partial \xi}{\partial w} 
\frac{\partial w}{\partial z}
\end{equation}
to arrive at:
\begin{equation}\label{eq:exiey}
E_x - i E_y = -\frac{\phi_0 b}{K(k)}\frac{1}{\sqrt{(z^2-a^2)(z^2-b^2)}}
\end{equation}

\section{Energy} \label{section:energy}

To calculate the electric field energy residing in a two-dimensional volume within the coplanar design (Fig. \ref{fig:padconform}a), we note that the electric field is merely the gradient of the potential $\phi(z)$.  Therefore, a solution to the volume integral
\begin{equation} \label{eq:green1}
\int_{V} {|\vec{E}|^2 dV} = \int_{V} {\nabla \phi \cdot \nabla \phi dV}
\end {equation}
can be simplified to a surface integral using Green's first identity if $\phi$ satisfies Laplace's equation $\nabla^2 \phi = 0$
\begin{equation} \label{eq:green2}
\int_{V} {\nabla \phi \cdot \nabla \phi dV} = -\oint_{S_i}{\phi \vec{E} \cdot \hat{n} dS}
\end{equation}
where $\hat{n}$ represents the outward unit normal vector along the surface.  We can use (\ref{eq:exiey}) to determine not only the electric field along this contour but also the potential since $\phi(z_2) - \phi(z_1) = -\int_{z_1}^{z_2}{\vec{E} \cdot dl}$ where $l$ corresponds to the path difference between arbitrary points $z_1$ and $z_2$ (see Appendix \ref{section:appcpc}).  By transforming the calculation of the electric field energy into an integration over a closed contour $S_i$ which does not need to contain the singularities at the edges of the metallization ($x = \pm a, \pm b$), (\ref{eq:green2}) will converge.  Thus, the electric field energy within an arbitrary volume with relative dielectric constant $\epsilon_r$ can be represented by:
\begin{equation}\label{eq:udef}
U_i = -\frac{\epsilon_0 \epsilon_r}{2}  \oint_{S_i}{\phi \vec{E} \cdot \hat{n} dS}
\end{equation}
where the subscript $i$ refers to a particular domain residing within the substrate.  For example, $U_{SM}$ and $U_{SA}$ refer to the energy in regions extending a thickness $\delta$ below the $SM$ and $SA$ interfaces, respectively, $U_{sub}$ refers to that contained within a thickness $\delta$ extending below the entire substrate top surface and $U_{tot}$ 
the energy of the system.  Note that for two-dimensional domains, the line integral in (\ref{eq:udef}) is evaluated about a clockwise loop.

\begin{figure}[htbp!]
		 \centering
	\includegraphics[width=0.35\textwidth]{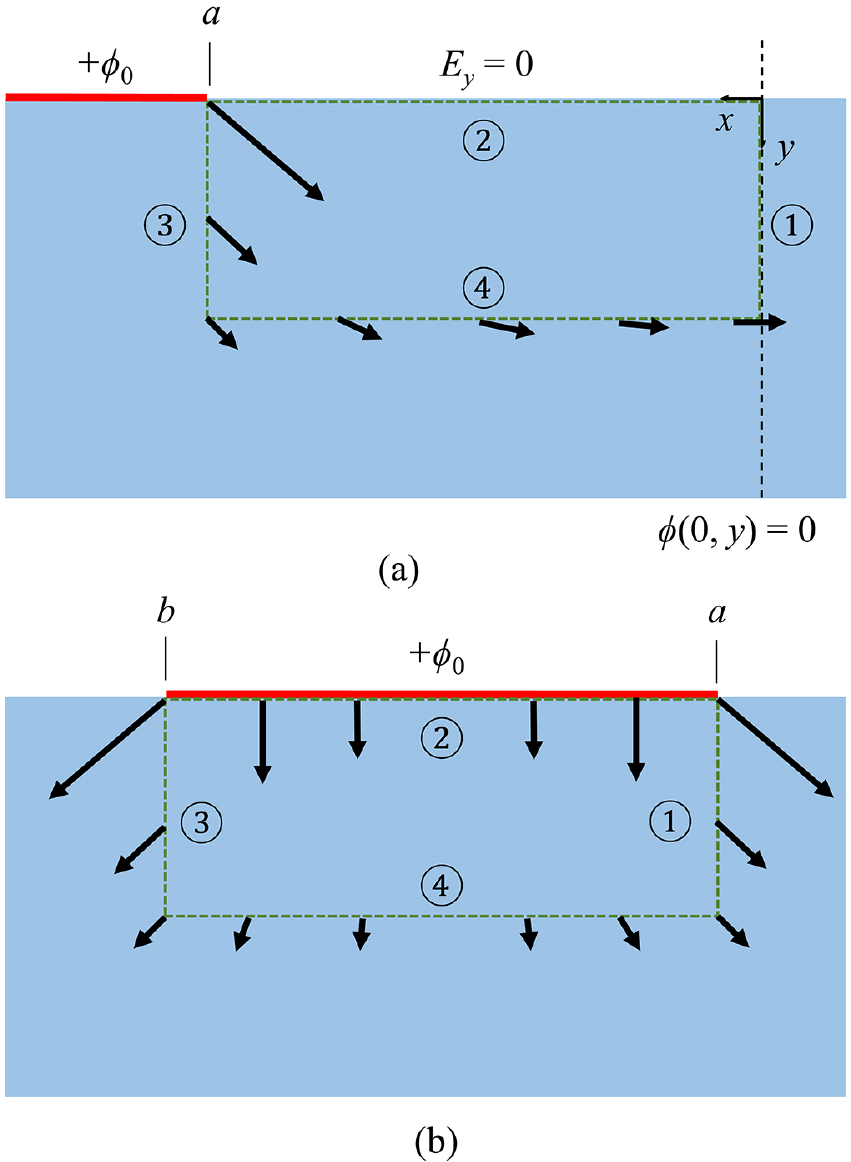}
		 \caption{(a) Calculation of electric field energy, U, within a rectangular region between the capacitor paddles.  The contribution corresponding to the y-axis (labelled 1) is zero because the potential is zero and the contribution along the x-axis (labelled 2) is also zero because the component of the electric field normal to the surface is zero.  Only surfaces 3 and 4 must be evaluated using  (\ref{eq:udef}). (b) Calculation of U for a rectangular region directly under one paddle, corresponding to half of $U_{SM}$, where all four surfaces contribute to the integral.}
		 \label{fig:surfint}
\end{figure}

Fig.'s \ref{fig:surfint}a and \ref{fig:surfint}b depict the methodology for calculating $U$ in a rectangular region within the substrate between the capacitor paddles and directly underneath a paddle, respectively.  
For the surface of a rectangular region aligned with the $x$ and $y$ axes,
$\vec{E} \cdot \hat{n}$ simply refers to the corresponding component of $\vec{E}$ normal to that surface.  The boundary conditions of $\phi$ allow us to further simplify the integral by noting that the potential is constant under the metallization ($\phi(x,0) = +\phi_{0}$ for $a \leq x \leq b$ and $-\phi_{0}$ for $-b \leq x \leq -a$), the potential field is antisymmetric with respect to $x$ $(\phi(0,y) = 0)$ and is zero at $\infty$.  These boundary conditions also allow us to directly calculate the total electric field energy, $U_{tot}$, of the system within Fig. \ref{fig:padconform}a by defining two volumes: one that encompasses the entire semi-infinite substrate and one enclosing the semi-infinite vacuum, and noting that the only contribution to (\ref{eq:udef}) comes from the surfaces adjacent to the electrodes:
\begin{equation}\label{eq:utotdef}
U_{tot} = \epsilon_0 \left( \epsilon_{sub} + 1 \right) \left(\phi_0 \right)^2 \frac{K(k')}{K(k)}
\end{equation}
Because the units in (\ref{eq:utotdef}) are Joules / meter due to the two-dimensional analysis of the electric fields,
the total energy can be determined by multiplying $U_{tot}$ by the length of the capacitor, $l_0$.  In the case of interdigitated
capacitor paddles, $l_0$ represents an effective length based on the aspect ratio of the paddles.  Although the paddle corners are neglected in the calculation of (\ref{eq:utotdef}), their contribution to $U_{tot}$, which is proportional to
the overall capacitance, is estimated to be less than $2 \%$  \cite{Alley70}.

\begin{figure}[htbp!]
		 \centering
	\includegraphics[width=0.4\textwidth]{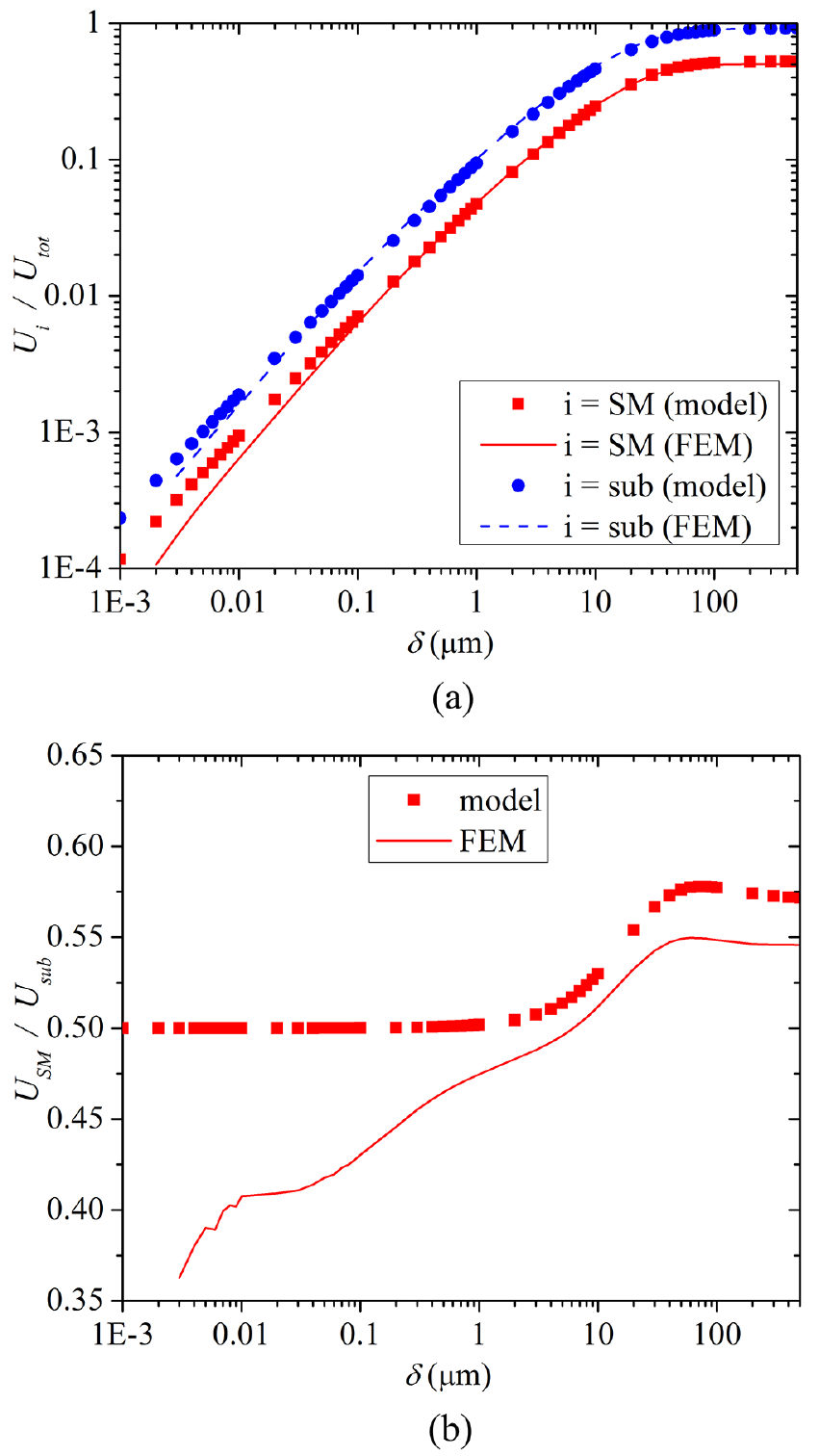}
		 \caption{(a) Comparison of electric field energy per unit length within a volume of thickness $\delta$ from the top surface of a dielectric substrate $\left(U_{sub}\right)$ or only under the substrate-to-metal regions $\left(U_{SM}\right)$ normalized by the total electric field energy $\left(U_{tot}\right)$, as calculated by FEM and by the analytical model (\ref{eq:udef}) and (\ref{eq:utotdef}). The paddle dimensions are $a$ = 10 $\mu$m and $b$ = 70 $\mu$m and substrate relative dielectric constant is 11.45. (b) Ratio of $U_{SM}$ to $U_{sub}$ from the analytical model and from FEM simulations. }
		 \label{fig:uivsutot}
\end{figure}

Fig. \ref{fig:uivsutot}a depicts $U_{sub}$ and $U_{SM}$ normalized by $U_{tot}$ as calculated by the surface integral model using Mathematica (Wolfram Research, Inc., Champaign, IL, USA) and by FEM using HFSS (Ansys, Inc. Canonsburg, PA, USA), where the paddle dimensions are $a$ = 10 $\mu$m and $b$ = 70 $\mu$m and substrate relative dielectric constant is 11.45.  The two methods diverge in their calculation of $U_{sub} / U_{tot}$ for values of $\delta$ less than 0.1 $\mu$m but both approach
$\epsilon_{sub} / \left( \epsilon_{sub}+1 \right) \approx 0.92$ at large $\delta$.  A similar discrepancy in
$U_{SM} / U_{tot}$ is observed in Fig. \ref{fig:uivsutot}a between the two methods for small values of $\delta$.  As can be seen in Fig. \ref{fig:uivsutot}b, the ratio of $U_{SM}$ to $U_{sub}$ is predicted to be 0.5 for $\delta \leq 1 \mu$m by the analytical model and increases slightly for larger values of $\delta$, with the remainder corresponding to the electric field energy under the free surface of the substrate, $U_{SA}$.  It is clear that for thin volumes the analytical model gives a better representation of the electric field energy than the finite element method.  The reasons for the observed differences between the two models involve the singularities in the electric field intensity at the metallization corners, and the difficulty in accurately catpuring their effects when discretizing domains in which large disparities between $\delta$ and the lateral dimensions of the paddles exist \cite{Wenner11}, \cite{Wang15}, \cite{Gambetta16}.  These complications are exacerbated in regions that are adjacent to such singularities and possess thicknesses less than 10 nm, both of which are relevant to the calculation of surface participation.

\section{Participation}  \label{section:partbc}

The approach described in Section \ref{section:energy} is applicable to calculate the electric field energy in any prescribed volume.  However, to calculate the volume participation due to thin layers with dielectric constants that differ from that of the substrate, we can utilize the matching of boundary conditions at the interface between the contamination layer and the substrate \cite{Wenner11},\cite{Sandberg12},\cite{Gambetta16},\cite{Dial16}:
\begin{eqnarray}
E_x^{c}  &=& E_x^{i}   \label{eq:eqn1} \\
\epsilon_{c} E_y^{c}  &=& \epsilon_{i} E_y^{i} \label{eq:eqn2}
\end{eqnarray}
where the superscript $c$ refers to the hypothetical contamination layer and $i$ the actual dielectric material present in the simulation (e.g: silicon for the substrate surfaces or vacuum for the free surfaces).
Within a region close to the top surface of the substrate $(y \ll a)$, the electrostatics dictate that underneath the electrodes $E_y^{c} \approx \frac{\epsilon_i}{\epsilon_c} E_y^{i}$ represents the dominant contribution to the electric field whereas  $E_x^{c} \approx E_x^{i}$ dominates in the regions without metallization.  In this way, we can approximate the participation in contamination layers below the paddle metallization $(U_{SM})$, above the paddle metallization $(U_{MA})$ and along the free substrate surface $(U_{SA})$ in the following manner:
\begin{eqnarray}
P_{SM} & \approx & \left( \frac{\epsilon_{sub}}{\epsilon_{c:SM}} \right)^2 \frac{U_{SM}^c}{U_{tot}} \label{eq:psm0} \\
P_{SA} & \approx & \frac{U_{SA}^c}{U_{tot}} \label{eq:psaeq0} \\
P_{MA} & \approx &  \left( \frac{1}{\epsilon_{c:MA}} \right)^2 \frac{U_{SM}^c}{U_{tot}} \label{eq:pma0}
\end{eqnarray}
where $U_i^c$ refers to the surface integral in (\ref{eq:udef}) over the corresponding volumes with relative dielectric constant $\epsilon_{c:i}$.  Note that it is assumed that the contribution of the electric field energy within the contamination layers is much smaller than that of the entire system.  From (\ref{eq:udef}), we can also express $U_i^c$ as
$\left( U_i \epsilon_{c:i} \right) / \epsilon_r$ so that (\ref{eq:psm0}) to (\ref{eq:pma0}) can be simplified to form:
\begin{eqnarray}
P_{SM} & \approx & \left( \frac{\epsilon_{sub}}{\epsilon_{c:SM}} \right) \frac{U_{SM}}{U_{tot}} \label{eq:psm} \\
P_{SA} & \approx & \left( \frac{\epsilon_{c:SA}}{\epsilon_{sub}} \right) \frac{U_{SA}}{U_{tot}} \label{eq:psaeq} \\
P_{MA} & \approx & \left( \frac{1}{\epsilon_{c:MA} \epsilon_{sub}} \right) \frac{U_{SM}}{U_{tot}} \label{eq:pma}
\end{eqnarray}
where the following relation, $U_{MA} = U_{SM} / \epsilon_{sub}$, was incorporated into (\ref{eq:pma}) due to the symmetry of the electric field distributions above and below the metallization. Because we can infer that  $U_{SM} \approx U_{SA}$ for values of $\delta / a < 0.1$ from Fig. \ref{fig:uivsutot}b, we can establish approximate ratios of the various surface participation components.
\begin{eqnarray}
\frac{P_{SM}}{P_{SA}} & \approx & \frac{ \left( \epsilon_{sub} \right)^2}{\epsilon_{c:SM} \epsilon_{c:SA}}
\label{eq:psmpsa} \\
\frac{P_{MA}}{P_{SM}} & \approx & \frac{ \epsilon_{c:SM}}{\epsilon_{c:MA} \left( \epsilon_{sub} \right)^2}
\label{eq:pmasm}
\end{eqnarray}
Equations (\ref{eq:psmpsa}) and (\ref{eq:pmasm}) demonstrate that the surface participation components can be treated as linearly dependent for small values of $\delta / a$ \cite{Wang15}, \cite{Calusine18}.

\begin{figure}[htbp!]
		 \centering
	\includegraphics[width=0.45\textwidth]{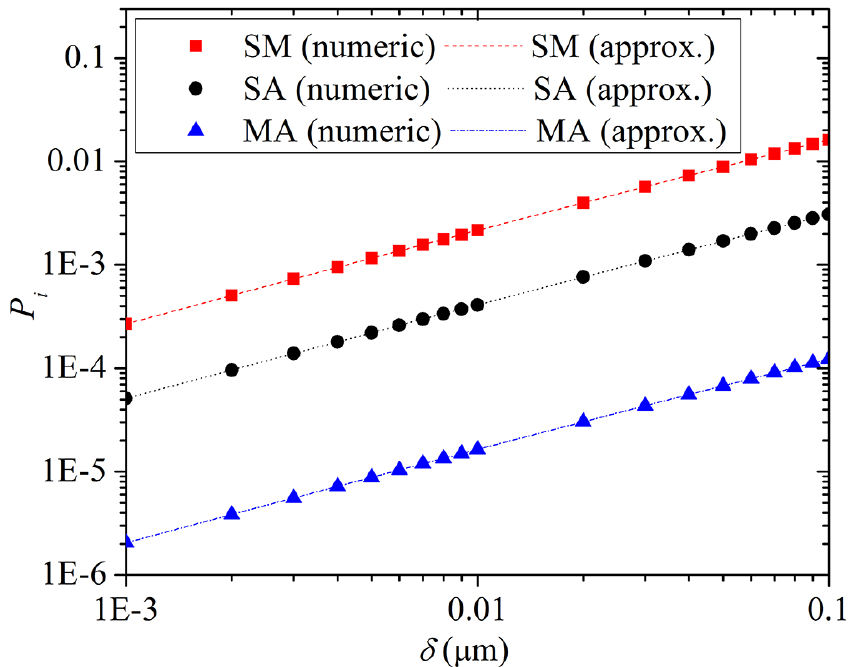}
		 \caption{Comparison of substrate-to-metal (SM), substrate-to-air (SA) and metal-to-air (MA) participation values as a function of contamination layer thickness, $\delta$, all with relative dielectric constant $\epsilon_c = 5.0$. 
The symbols correspond to values calculated numerically using (\ref{eq:psm}) to (\ref{eq:pma}) and the dashed lines the approximate formulation (\ref{eq:psa}) combined with (\ref{eq:psmpsa}), (\ref{eq:pmasm}). The paddle dimensions are $a$ = 10 $\mu$m and $b$ = 70 $\mu$m and substrate relative dielectric constant is 11.45.}
		 \label{fig:piall}
\end{figure}

The symbols depicted in Fig. \ref{fig:piall} represent the participation values calculated numerically using the surface integral approach (\ref{eq:psm}) to (\ref{eq:pma}) as a function of contamination layer thickness, $\delta$, for a paddle design with $a$ = 10 $\mu$m and $b$ = 70 $\mu$m, $\epsilon_{sub} = 11.45$ and equal relative dielectric constants, $\epsilon_c = 5.0$, among all of the contamination layers.  It is clear that the SM participation represents the largest contribution and that the MA participation is over two orders of magnitude less: $(1/\epsilon_{sub})^2$ from (\ref{eq:pmasm}).  The finding of small $P_{MA}$ values relative to $P_{SM}$ or $P_{SA}$ is consistent with the calculations of \cite{Wenner11} and \cite{Wang15}.  However, the ratio of $P_{SM}$ to $P_{SA}$ of approximately 5.2, $ (\epsilon_{sub} / \epsilon_c)^2$ from 
(\ref{eq:psmpsa}), differs from \cite{Wenner11} and \cite{Wang15} primarily due to the values of relative dielectric constants assumed among the various simulations.  Although the true compositions of the surface contamination layers are not known, a value for $\epsilon_c$ of 5.0 would be representative of organic residue or silicon oxide, which we believe to remain on the top surface of silicon substrates.  Other simulations have used $\epsilon_c$ values of 10.0 which were equal to the corresponding substrate dielectric constants \cite{Wenner11}, \cite{Wang15}.  From (\ref{eq:psm}) and (\ref{eq:psaeq}), we see that identical values of $\epsilon_c$ and $\epsilon_{sub}$ should produce ratios of $P_{SM}$ to $P_{SA}$ of approximately 1.

\section{Participation approximation} \label{section:approx}

These results can be compared to a much simpler representation based on surface participation as follows.  We can directly integrate (\ref{eq:exiey}) with respect to $x$ at a specific value of $y$ where we restrict its evaluation to the region
$y \ll a$.  As derived in Appendix \ref{section:appapprox}, $r_i (y / a)$ within a contamination layer with relative dielectric constant $\epsilon_c$ and underneath the SA interface can be represented as a logarimthic function with respect to the ratio $y/a$:
\begin{eqnarray}
& & r_{SA}^c \left(\frac{y}{a} \right) = \frac{\epsilon_c}{(\epsilon_{sub} + 1)} \frac{1}{2a(1-k)K'(k) K(k)} \nonumber \\
& & \cdot \left\{ \ln \left[ 4 \left( \frac{1-k}{1+k} \right)  \right] - \frac{k \ln(k)}{(1+k)} - \ln \left( \frac{y}{a} \right) \right \} \label{eq:rsa}
\end{eqnarray}
It is straightforward to integrate this equation with respect to $y$ to derive an approximate formulation of the electric field energy within a volume of thickness $\delta$ from the top surface of the substrate:
\begin{eqnarray}
P_{SA} \left(\frac{\delta}{a} \right) \approx \int_0^\delta  r_{SA}^c \left(\frac{y}{a} \right) dy \nonumber \\
 = \frac{\epsilon_c}{(\epsilon_{sub} + 1)}\frac{1}{2(1-k)K'(k) K(k)} \cdot \nonumber \\
  \left( \frac{\delta}{a} \right) \left\{ \ln \left[ 4 \left( \frac{1-k}{1+k} \right)  \right] - \frac{k \ln(k)}{(1+k)}+1
 - \ln \left( \frac{\delta}{a} \right) \right \} \label{eq:psa}
\end{eqnarray}
Participation under the various surfaces can then be calculated using the same matching procedure with respect to the electric fields with and without a contamination layer. For example, participation associated with the SM interface is calculated by combining (\ref{eq:psmpsa}) with (\ref{eq:psa}):
\begin{eqnarray} \label{eq:psmapp}
P_{SM} \left(\frac{\delta}{a} \right) \approx \frac{\epsilon_{sub}^2}{\epsilon_c \left( \epsilon_{sub}+1 \right)}
\frac{1}{2(1-k)K'(k) K(k)} \cdot \nonumber \\
 \left( \frac{\delta}{a} \right) \left\{ \ln \left[ 4 \left( \frac{1-k}{1+k} \right)  \right] - \frac{k \ln(k)}{(1+k)}+1
 - \ln \left( \frac{\delta}{a} \right) \right \}
\end{eqnarray}
As seen in Fig. \ref{fig:piall}, the dashed lines corresponding to (\ref{eq:psa}) combined with (\ref{eq:psmpsa}) and (\ref{eq:pmasm}) closely match the numerically calculated values using the surface integral formulation.  Because these curves appear linear in Fig. \ref{fig:piall}, a power law fitting can be performed with an exponent value of approximately 0.88, also arrived at in \cite{Gao08} for CPW structures.  However, as shown in (\ref{eq:rsa}) and (\ref{eq:psa}), participation is represented by the integral of a function with a logarithmic dependence on $y / a$ which accounts for the deviation from an exponent of 1 and is not due to effects of finite thickness of the metallization \cite{Gao08}.  In fact, the value of this exponent will vary depending on the range in
$\delta / a$ considered in such a fitting.

\begin{figure}[htbp!]
		 \centering
	\includegraphics[width=0.45\textwidth]{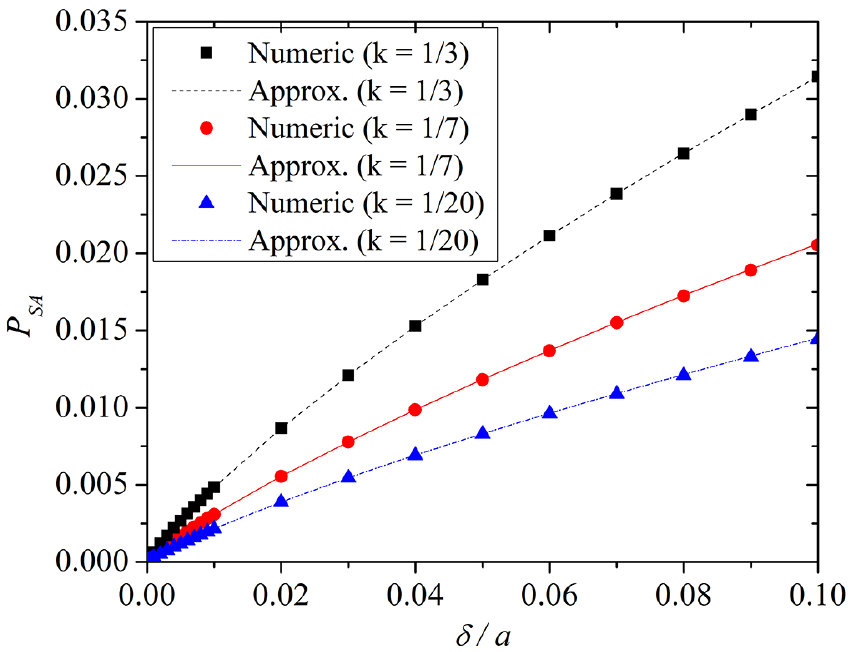}
		 \caption{Calculation of substrate-to-air (SA) participation values as a function of the ratio of contamination layer thickness, $\delta$, to $a$ with relative dielectric constant $\epsilon_c = 5.0$.}
		 \label{fig:psavsb}
\end{figure}

Based on (\ref{eq:psa}), universal curves of participation can now be generated for a given value of $k$ based on the capacitor design dimensions.  Fig. \ref{fig:psavsb} depicts SA participation as a function of the dimensionless ratio of the contamination layer thickness to the paddle spacing, $\delta / a$, for several values of $b$.  As $k$ increases from 1/3, corresponding to equal paddle width and spacing values, to 1/7, where the paddle widths are three times as large as their spacing, $P_{SA}$  only decreases by approximately 38\%, demonstrating a weak dependence on $k$.  As shown in Fig. \ref{fig:psavsb}, the approximate formulation provides a close match (less than 0.5\% difference for $\delta / a$ = 0.1) to the participation values calculated by numerical surface integration.

\section{Coplanar waveguides} \label{section:cpw}
\begin{figure}[htbp!]
		 \centering
	\includegraphics[width=0.35\textwidth]{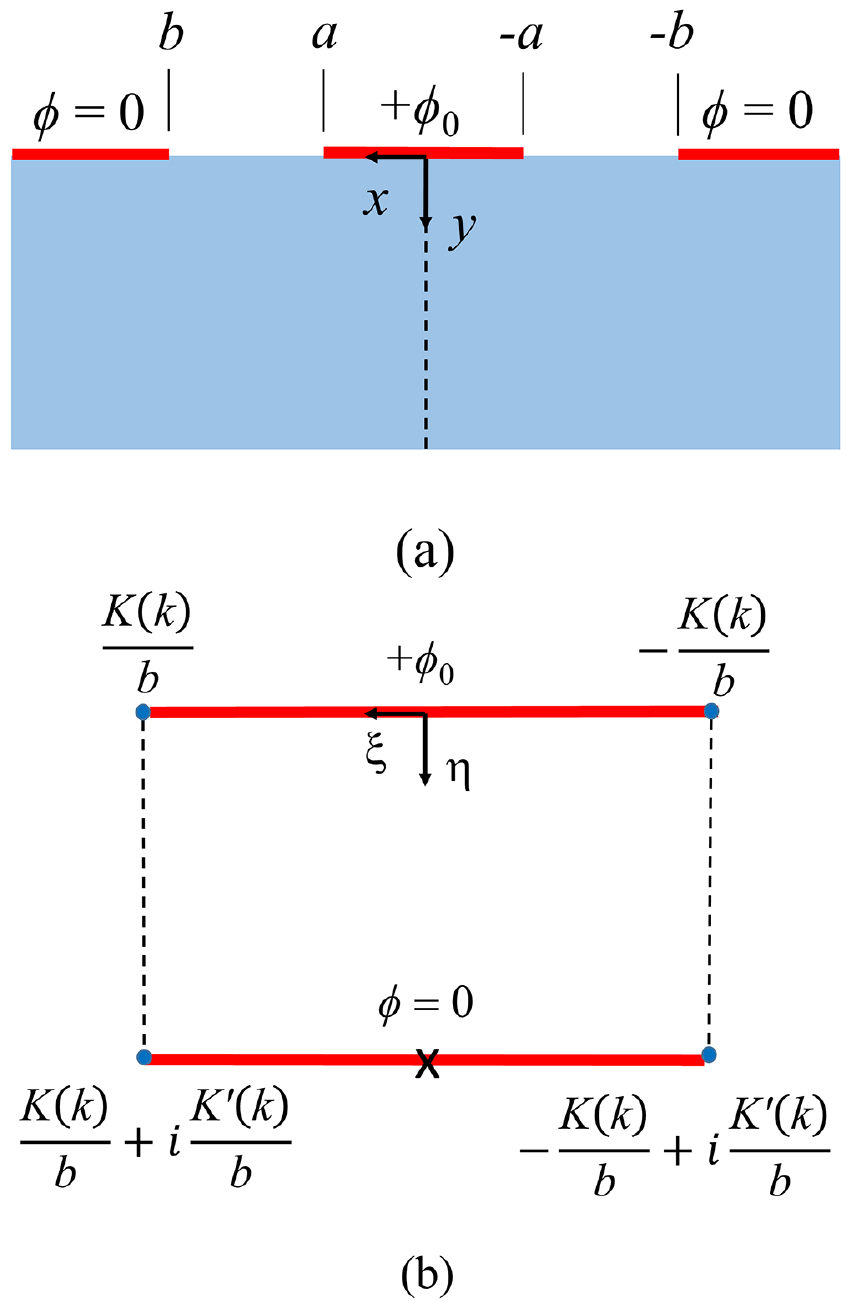}
		 \caption{(a) Cross-sectional schematic of a two-dimensional CPW metallization deposited on a semi-infinite substrate. (b) Transformed geometry through conformal mapping from the $x - y$ plane to the $\xi-\eta$ plane using (\ref{eq:eqcon1}).  The x on the bottom metallization denotes the point corresponding to $-\infty$ and $+\infty$ along the $x$-axis from Fig. \ref{fig:figcpw}a.}
		 \label{fig:figcpw}
\end{figure}

The metallization associated with a CPW design, located on the top surface of a dielectric substrate, is inverted relative to that of Fig. \ref{fig:padconform}a.  As shown in Fig. \ref{fig:figcpw}a, a potential of $+\phi_0$ is applied to the center conductor, which resides between $x = -a$ and $x = a$, and the groundplane (at zero potential) spans the distance from $b$ to $\infty$ and from $-b$ to $-\infty$.  
Although the same conformal mapping from the $z$-plane to $w$-plane (\ref{eq:eqcon1}) can be applied, the electric field corresponding to the even mode of the CPW is aligned parallel to the $\eta$ axis, taking the form:
\begin{equation} \label{eq:ecpww}
E_\xi - i E_\eta = -i \frac{\phi_0 b}{K'(k)}
\end{equation}
with the following complex potential distribution:
\begin{equation} \label{eq:phicpww}
\phi(\eta) = \phi_0 \left[1- \frac{b}{K'(k)} \eta \right]
\end{equation}
The transformed electric field distribution in the $z$-plane can be represented as:
\begin{eqnarray} \label{eq:exieycpw}
E_x - i E_y & = &  -\frac{\partial \phi \left( \eta \right)}{\partial \eta} \frac{\partial \eta}{\partial w} 
\frac{\partial w}{\partial z} \nonumber \\
& = & -i \frac{\phi_0 b}{K'(k)}\frac{1}{\sqrt{(z^2-a^2)(z^2-b^2)}} \label{eq:exieycpw}
\end{eqnarray}
The same procedure for calculating the electric field energy through surface integration can be employed, where the electric field distribution and corresponding potential come from (\ref{eq:exieycpw}).  Note that the total energy per unit length of the CPW has a different value than in the case of the coplanar capacitors (\ref{eq:utotdef}):
\begin{equation}\label{eq:utotcpw}
U_{tot}^{CPW} = \epsilon_0 \left( \epsilon_{sub} + 1 \right) \left(\phi_0 \right)^2 \frac{K(k)}{K(k')}
\end{equation}

Participation values for a CPW structure possess the same form as those shown in (\ref{eq:psm}) to (\ref{eq:pma}), where the surface integral is performed over the appropriate regions (e.g: $ a \leq |x| \leq b $ for SA). It is of interest to note that the analytical approximation (\ref{eq:psa}) holds for both coplanar capacitors and waveguides, as shown in Appendix \ref{section:appapprox}.

\section{Comparison}

\begin{figure}[htbp!]
		 \centering
	\includegraphics[width=0.45\textwidth]{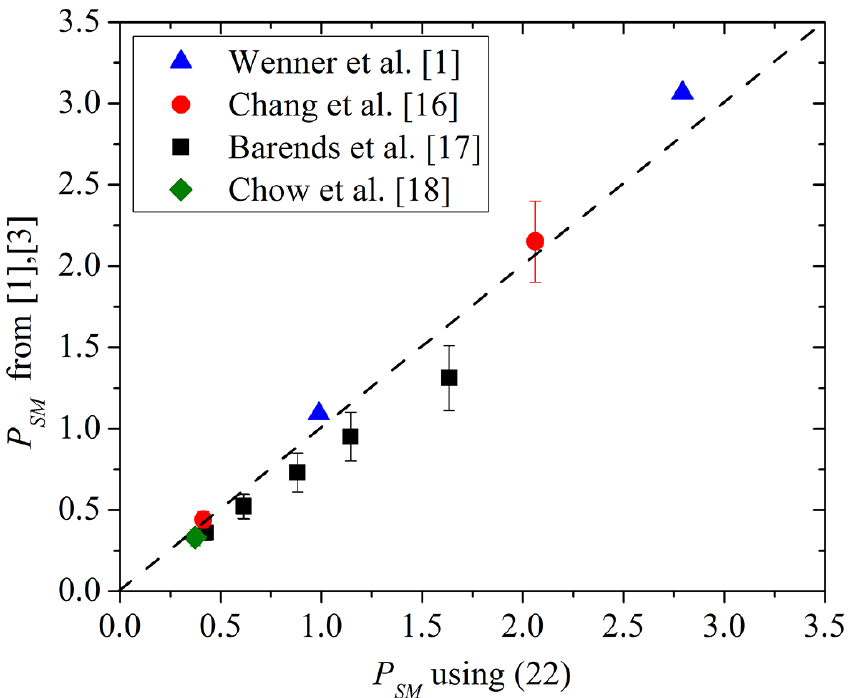}
		 \caption{Comparison of calculated surface participation values, $P_{SM}$, using (\ref{eq:psmapp}) to those predicted in \cite{Wenner11}, \cite{Wang15} for various CPW resonator and qubit designs.  The latter reference contains geometries listed in \cite{Chang13},\cite{Barends13},\cite{Chow12}.  The dashed line corresponds to a ratio of 1.}
		 \label{fig:comp}
\end{figure}

It is instructive to compare the method of calculating surface participation presented in Section \ref{section:approx}
to those simulated using different approaches in the previous literature.  Fig. \ref{fig:comp} depicts calculations of the SM surface participation based on the approximate formulation (\ref{eq:psmapp}) on the horizontal axis, and the corresponding values based on various untrenched, coplanar designs extracted from \cite{Wenner11} and \cite{Wang15} on the vertical axis.  The latter reference contains a survey of $P_{SM}$ values that were calculated for CPW resonators and qubits based on geometries reported in \cite{Chang13}, \cite{Barends13} and \cite{Chow12}.  For these simulations, both \cite{Wenner11} and \cite{Wang15} assume equal values for the material and geometric parameters ($\delta$ = 3 nm, $\epsilon_{c:SM}$ = 10) and $\epsilon_{sub}$ = 10 or 11.7 for structures fabricated on sapphire or silicon substrates, resepctively.  Qubits with interdigitated capacitors were modeled using a finger width of $2a$ and gap between fingers of $b-a$.  Likewise, CPW resonators possessed centerline conductor widths of $2a$ and gaps of $b-a$.  As shown in Fig. \ref{fig:comp}, the calculated values from (\ref{eq:psmapp}) replicate those reported in the prior literature, despite the assumptions of a two-dimensional design and metallization features with zero thickness.  There is a good match betweeen (\ref{eq:psmapp}) and the approach of \cite{Wenner11}, that employs a complex combination of FEM simulations and modifications to the magnitude of the electric fields near the metallizaton corners,
and to methods \cite{Wang15} that stitch together multiple FEM models at different length scales.

\section{Application to a transmon qubit}

\begin{figure}[htbp!]
		 \centering
	\includegraphics[width=0.45\textwidth]{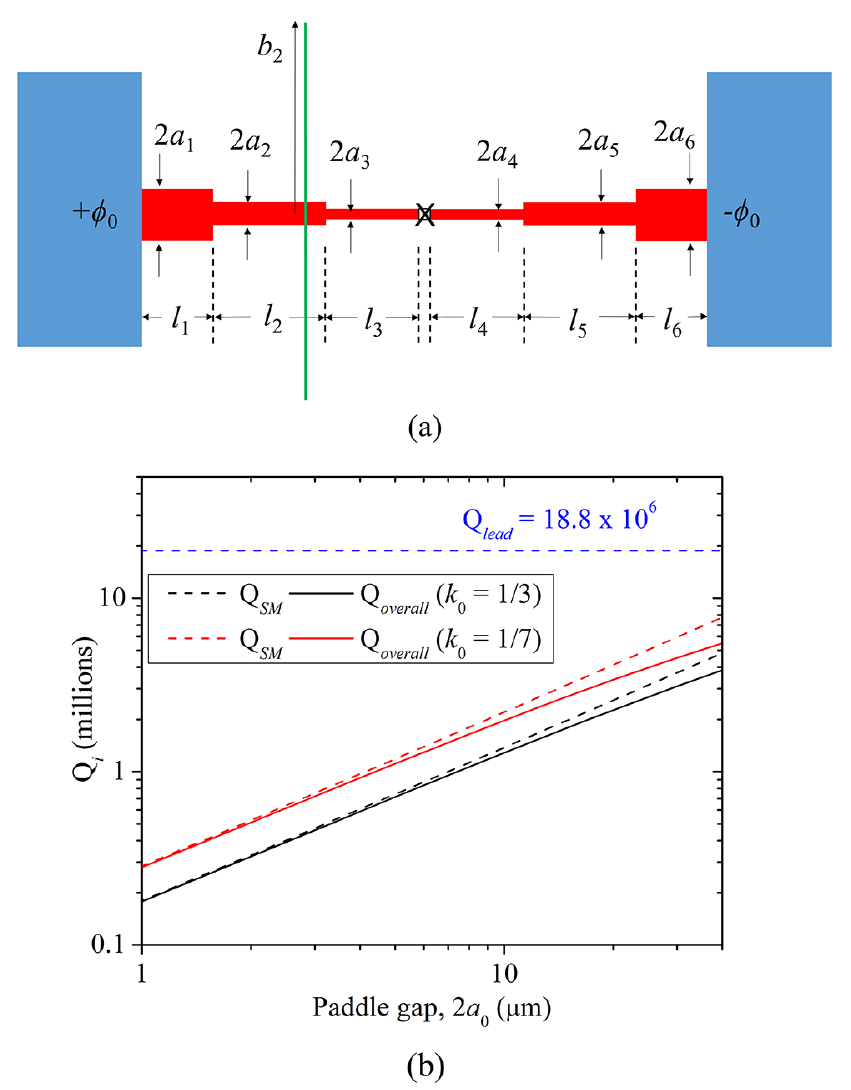}
		 \caption{(a) Top-down schematic of the portion of a qubit design containing the junction, leads and adjacent capacitor paddle edges.  The coplanar paddles are assumed to be at an electrical potential of $+\phi_0$ and $-\phi_0$, respectively, with a gap size of $2a_0$.  The junction leads are approximated by 6 rectangular segments, each with width $2a_i$, distance between
the centerline of the leads and the edges of the grounding plane (not shown) $\pm b_i$ and length $l_i$.
Calculation of the surface participation of the leads
is accomplished through a piecewise integration along the individual segments using the CPW approximation
(Fig. \ref{fig:figcpw}a), where the cross-section is aligned with the green solid line. (b) Calculated Q values due to SM participation
 (tan$\left(\delta_{SM}\right) = 4.8 \cdot 10^{-4}$),
junction lead participation (tan$\left(\delta_{lead}\right) = 4.8 \cdot 10^{-3}$) and overall Q based on these contributions
as a function of capacitor paddle gap ($2a_0$) for two different $k_0$ values.  The contamination layer thickness is constant
($\delta = 2$ nm with a relative dielectric constant $\epsilon_c = 5.0$).}
		 \label{fig:qtotab}
\end{figure}

Simulations of qubit architectures often require the combination of electric field distributions conducted over a large scale, comparable in size to the substrate, with finer calculations local to the junction region.  Because of the large disparity in length scales associated with these two regimes, it is often impractical to use one simulation to account for all effects, and can require schemes that stitch together two different FEM models \cite{Wang15}.  However, such approaches are still susceptible to errors arising from singularities in the electric field near metallization corners as well as those emanating from overlapping two different solutions over an arbitrarily defined transition region.  By using the analytic models presented in the previous sections, we can avoid these issues and employ the quasi-static approximation to match the electric potential, for example, at the intersection of the shunting capacitor paddles and the junction leads.  The overall surface participation represents a linear superposition of the
effects from these regions.  For transmon qubit designs that incorporate a large shunting capacitance (such as those in \cite{Gambetta16}), the large-scale electric field distributions will be dominated by these coplanar capacitors, which can be interdigitated or monolithic in shape.  A model geometry is depicted in the top-down schematic of Fig. \ref{fig:qtotab}a, which contains the qubit junction, leads and the edges of the adjacent capacitor paddles.  In this example, the geometry of the junction leads is assumed to be composed of 6 individual coplanar sections, each with a width of $2a_i$, length $l_i$, and distance
from the centerline of the leads to the grounding plane $b_i$. 
A symmetric design with respect to the junction position is considered, consisting of a 2 $\mu$m wide, 7$\mu$m long section
(\#1 and \#6), an 0.5 $\mu$m wide, 2 $\mu$m long section (\#2 and \#5) and a 0.1 $\mu$m wide, 1 $\mu$m long section
(\#3 and \#4) adjacent to the junction.  We can use the results of Sections \ref{section:energy} to \ref{section:cpw}
to calculate the corresponding surface participation.

Let us assume that SM participation represents the dominant surface loss mechanism \cite{Wisbey10}, where a 2 nm thick contamination layer with relative dielectric constant $\epsilon_c = 5.0$ and loss tangent of $4.8 \cdot 10^{-4}$ (see Appendix \ref{section:applosstan}) resides between the metallization and the silicon substrate.  In the quasi-static approximation shown in Fig. \ref{fig:qtotab}a, a positive and negative potential of magnitude $\phi_0$ exists on either capacitor paddle, similar to that shown in Fig. \ref{fig:padconform}a.  The value of $\phi_0$ does not need to be known to determine surface participation since $\phi_0$ cancels from the numerator and denominator in (\ref{eq:psm}) to (\ref{eq:pma}) for the analytical calculation or (\ref{eq:psmapp}) for the analytical approximation.  The corresponding quality factor, Q, is equal to the inverse of the product of the SM participation and the loss tangent of the contamination layer, tan$(\delta_{SM})$.  The dotted lines in Fig. \ref{fig:qtotab}b depict the Q values predicted as a function of paddle gap, $2a_0$, for two different values of $k_0$ based solely on SM participation. 

The procedure outlined in Section \ref{section:cpw}, corresponding to a CPW design, can be extended to determine the contribution of the qubit junction leads to the overall participation. The green, solid vertical line in Fig. \ref{fig:qtotab}a corresponds to the cross-sectional plane represented in Fig. \ref{fig:figcpw}a where $2 a$ refers to the width of one section of the junction leads and $\pm b$ is the distance from the centerline of the leads to the grounding planes on either side of the qubit.  If we again assume that SM participation dominates surface loss, the surface participation associated with junction leads that vary in width along their length can be calculated by summing contributions from each individual section:
\begin{eqnarray} \label{eq:ptotleads}
P_{lead}^{SM} & \approx & \frac{\sum_{i=1}^n P_{i}^{SM} U_i l_i}{U_{tot} l_0} \nonumber \\
& = & \frac{ \epsilon_{0} \left( \epsilon_{sub}+1 \right) 
\left( \phi_0 \right)^2}{\frac{1}{2} C_q  \left( 2 \phi_0 \right)^2}
\sum_{i=1}^{n} P_{i}^{SM} \frac{K(k_i)}{K'(k_i)} l_i \nonumber \\
& \approx & \frac{\epsilon_{0} \left( \epsilon_{sub}+1 \right)}{2 C_q} K(0) \sum_{i=1}^{n} \frac{P_{i}^{SM} l_i}{K'(k_i)}
\end{eqnarray}
where the $i^{th}$ section possesses a length, $l_i$, width, $2 a_i$, and $k_i$ represents the ratio of the lead width to the distance between the edges of the grounding plane, $2 b_i = 2 b$.  $U_i$ corresponds to the total energy per unit length as determined from (\ref{eq:utotcpw}) and $P_i^{SM}$ can be calculated by the approximate formulation (\ref{eq:psmapp}). 
In the limit of small $k$, $K(0)$ approaches $\frac{\pi}{2}$ and can be moved outside of the summation in the numerator of (\ref{eq:ptotleads}).  The value of $2 \phi_0$ in the denominator arises from the fact that the capacitor paddles are at opposite polarities of magnitude $\phi_0$.  If we assume a contamination layer in the vicinity of the leads with the same thickness and relative dielectric constant but larger loss tangent, tan$\left(\delta_{lead}\right)$ of $4.8 \cdot 10^{-3}$, which is more representative of lift-off Al metallization (see Appendix \ref{section:applosstan}), then the Q value calculated only due to SM participation of the leads is $18.8 \cdot 10^{6}$.  This value is plotted in Fig. \ref{fig:qtotab}b along with the total Q, calculated by the sum of the reciprocal Q values due to SM capacitor paddle and SM lead participation, indicating that surface loss near the junction has a minor impact on the overall Q values in transmon qubit designs, particularly those that incorporate smaller shunting capacitor dimensions.  

\section{Conclusion}

In summary we present a two-dimensional, analytical formulation for electric field energy within prescribed volumes, based on a surface integration of the potential and electric field, which provides better accuracy at capturing the effects of the singular behavior of the electric fields near the metallization edges than finite element method models.  This approach has been applied to the calculation of participation at specific regions within coplanar capacitor and CPW designs, from which simple, closed-form expressions have been generated that approximate the participation of thin contamination layers.  The resulting participation values can be used to combine the effects of global electric field distributions from large-scale features with those local to the junction environment to arrive at overall quality factors of qubit designs.

\appendices
\section{Derviation of analytic formulae for electric field and potential in coplanar designs}\label{section:appcpc}

\begin{figure}[htbp!]
		 \centering
	\includegraphics[width=0.45\textwidth]{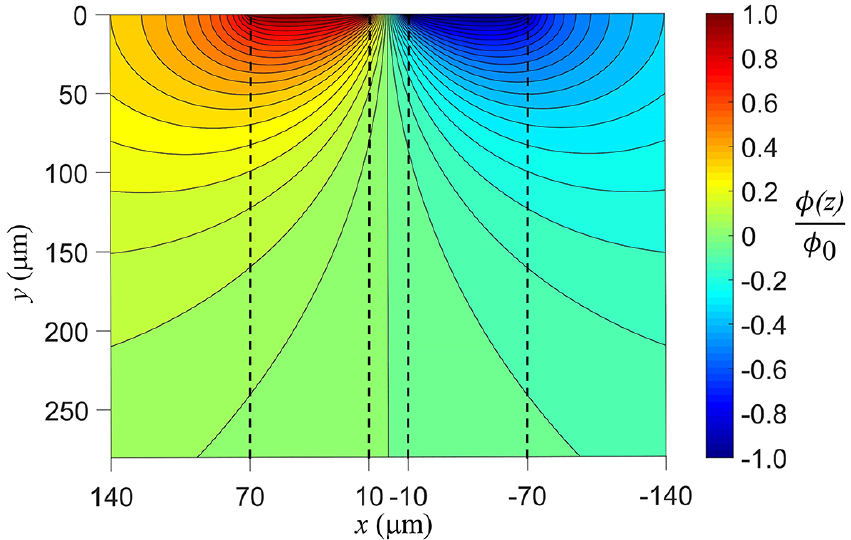}
		 \caption{Normalized electric potential $\phi(z) / \phi_0$ within the dielectric substrate as a function of position as calculated by (\ref{eq:phieqn}) for coplanar capacitor paddles where $a$ = 10 $\mu$m and $b$ = 70 $\mu$m.  Dashed vertical lines denote the position of the paddles located at the top surface of the substrate. }
		 \label{fig:phiwdashed}
\end{figure}

In the following appendix, we give the explicit forms of the electric potential and corresponding electric field distributions necessary to calculate the electric field energy within an arbitrary rectangular volume according to (\ref{eq:udef}). For a two-dimensional representation of a coplanar capacitor geometry, where the metallization, spanning from $a$ to $b$ and from -$a$ to -$b$, is assumed to be infinitely thin, as shown in Fig. \ref{fig:padconform}a, the potential takes the form $(y > 0)$:
\begin{equation}\label{eq:phieqn}
\phi (z) = \frac{\phi_0}{K(k)} \Re \left\{ F\left[ \sin^{-1} \left( \frac{z}{a} \right) , k \right] \right\}
\end{equation}
where $\Re$ refers to the real part of the expression, $z = x + i y$ and $F$ is the incomplete elliptic integral of the first kind.
Fig. \ref{fig:phiwdashed} depicts the potential distiribution, as normalized by the applied potential $\pm \phi_0$, through the underlying dielectric substrate.

The $x$ and $y$ components of the electric field are directly derived from (\ref{eq:exiey}).  However, their signs are affected by which branch cut is used in the complex $z$-plane.  For cases where $x^2 \leq y^2 + \left( a^2 + b^2 \right) / 2$:
\begin{eqnarray}
& & E_x (z) = - \frac{\phi_0 b}{K(k)} \Re \left[ \frac{1}{\sqrt{(z^2-a^2)(z^2-b^2)}} \right] \nonumber \\
& & E_y (z) =   \frac{\phi_0 b}{K(k)} \Im \left[ \frac{1}{\sqrt{(z^2-a^2)(z^2-b^2)}} \right]\label{eq:esmallx}
\end{eqnarray}
where $\Im$ refers to the imaginary part of the expression, or
\begin{eqnarray}
& & E_x (z) =  \frac{\phi_0 b}{K(k)} \Re \left[ \frac{1}{\sqrt{(z^2-a^2)(z^2-b^2)}} \right] \nonumber \\
& & E_y (z) =  -\frac{\phi_0 b}{K(k)} \Im \left[ \frac{1}{\sqrt{(z^2-a^2)(z^2-b^2)}} \right]\label{eq:ebigx}
\end{eqnarray}
when  $x^2 > y^2 + \left( a^2 + b^2 \right) / 2$.

\begin{figure}[htbp!]
		 \centering
	\includegraphics[width=0.45\textwidth]{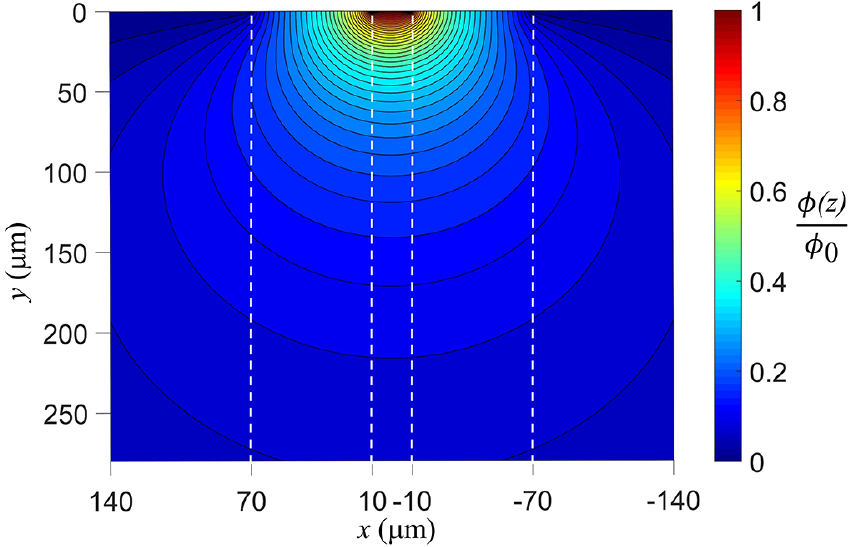}
		 \caption{Normalized electric potential $\phi(z) / \phi_0$ within the dielectric substrate as a function of position as calculated by (\ref{eq:phicpw}) for a coplanar waveguide with $a$ = 10 $\mu$m and $b$ = 70 $\mu$m.  Dashed vertical lines denote the position of the metallization edges located at the top surface of the substrate. }
		 \label{fig:cpwpot}
\end{figure}

For a CPW (Fig. \ref{fig:figcpw}a), where the center conductor is at a potential of $\phi_0$, the potential distribution takes a different form:
\begin{equation}\label{eq:phicpw}
\phi (z) = \phi_0 \left(1-\frac{1}{K'(k)} \Im \left\{ F\left[ \sin^{-1} \left( \frac{z}{a} \right) , k \right] \right\} \right)
\end{equation}
which is displayed in Fig. \ref{fig:cpwpot}.
The orthogonal components of the electric field can be represented by:
\begin{eqnarray}
& & E_x (z) =   \frac{\phi_0 b}{K'(k)} \Im \left[ \frac{1}{\sqrt{(z^2-a^2)(z^2-b^2)}} \right] \nonumber \\
& & E_y (z) =   \frac{\phi_0 b}{K'(k)} \Re \left[ \frac{1}{\sqrt{(z^2-a^2)(z^2-b^2)}} \right]\label{eq:ecpwsmallx}
\end{eqnarray}
for  $x^2 \leq y^2 + \left( a^2 + b^2 \right) / 2$ or
\begin{eqnarray}
& E_x (z) =  -\frac{\phi_0 b}{K'(k)} \Im \left[ \frac{1}{\sqrt{(z^2-a^2)(z^2-b^2)}} \right] \nonumber \\
& E_y (z) =  -\frac{\phi_0 b}{K'(k)} \Re \left[ \frac{1}{\sqrt{(z^2-a^2)(z^2-b^2)}} \right]\label{eq:ecpwbigx}
\end{eqnarray}
when  $x^2 > y^2 + \left( a^2 + b^2 \right) / 2$.

\section{Derivation of surface participation approximation formula}\label{section:appapprox}

The previous appendix describes the calculation of electric field energy within a prescribed rectilinear region
(\ref{eq:udef}), which holds for all volumes within a dielectric substrate but are solved numerically.  An analytical approximation can be generated for shallow depths relative to the in-plane dimensions of the metallization features.  Let us consider the description of the complex electric field (\ref{eq:exiey}) and determine the square of its norm, $|\vec{E}|^2$, as a function of position:
\begin{eqnarray}
|\vec{E}|^2  =  \Re[\vec{E}]^2 + \Im[\vec{E}]^2 = \left(\frac{\phi_{0} b}{K(k)} \right)^2  \cdot \nonumber \\
 \frac{1} {\sqrt{(x+a+iy)(x+a-iy)(x-a+iy)(x-a-iy)}} \cdot \nonumber \\
 \frac{1}{\sqrt{(x+b+iy)(x+b-iy)(x-b+iy)(x-b-iy)}} \label{eq:appb1}
\end{eqnarray}
which, in the case of $0 < y \ll a$ exhibits finite, local maxima near $x = \pm a$ and $x = \pm b$.  Equation (\ref{eq:appb1}) can be approximated using the auxiliary functions:
\begin{equation}\label{eq:thetaa}
\theta_a (x,y) = \left[\frac{\phi_{0} b}{K(k)}\right]^2\frac{1}{(b^2-x^2)(x+a)}\frac{1}{\sqrt{(x-a)^2+y^2}}
\end{equation}
near $x = a$ and:
\begin{equation}\label{eq:thetab}
\theta_b (x,y) = \left[\frac{\phi_{0} b}{K(k)}\right]^2\frac{1}{(x^2-a^2)(x+b)}\frac{1}{\sqrt{(x-b)^2+y^2}}
\end{equation}
near $x = b$.
Let $\Theta$ refer to the integration of (\ref{eq:thetaa}) as a function of $x$ from $0$ to $a$ and (\ref{eq:thetab})
from $x = b$ to $\infty$:
\begin{equation} \label{eq:psidef}
\Theta(y) =  \left[ \int_0^a \theta_a (x,y) dx + \int_b^\infty \theta_b (x,y) dx \right]
\end{equation}
corresponding to the SA regions.  In the limit of $y \ll a$, (\ref{eq:psidef}) can be simplified to form:
\begin{eqnarray}
& & \Theta \left(\frac{y}{a} \right) \approx  \left[ \frac{\phi_{0}}{K(k)} \right]^2  \frac{1}{[2a(1-(a/b)^2)]} \nonumber \\
& & \cdot   \left\{ \left(1+ \frac{a}{b} \right) \left( \ln \left[ 4 \left( \frac{b-a}{b+a} \right) \right]
-\ln \left[ \frac{y}{a} \right] \right) + \frac{a}{b} \ln \left[ \frac{b}{a} \right] \right\} \nonumber \\
& &=  \left[ \frac{\phi_{0}}{K(k)} \right]^2 \frac{1}{2 a (1-k)} \nonumber \\
& & \cdot \left\{ \ln \left[ 4 \left( \frac{1-k}{1+k} \right) \right] - \ln \left[ \frac{y}{a} \right] - \frac{k \ln(k)}{1+k} \right\} \label{eq:saenergy}
\end{eqnarray}
Note that for small values of $y$, an integration of (\ref{eq:appb1}) over SA is equal to that over the SM surface.

Equation (\ref{eq:saenergy}) allows us to represent surface participation at a specific depth, $y$ within a contamination layer
with relative dielectric constant, $\epsilon_c$ as:
\begin{equation} \label{eq:rsadef}
r_{SA} \left( \frac{y}{a} \right) = \frac{\epsilon_0 \epsilon_c}{U_{tot}} \Theta \left( \frac{y}{a} \right)
\end{equation}
which can be simplified, using (\ref{eq:utotdef}), to form:
\begin{equation}\label{eq:apprsa}
r_{SA}\left(\frac{y}{a}\right) \approx \frac{\epsilon_c}{(\epsilon_{sub} + 1)} \left[C_1 + C_2 \ln \left(\frac{y}{a} \right) \right] \frac{1}{2 a K(k') K(k)}
\end{equation}
where
\begin{eqnarray}
C_1 & = &  \frac{1}{(1-k)} \left\{ \ln \left[ 4 \left( \frac{1-k}{1+k} \right) \right] - \frac{k \ln(k)}{1+k} \right\} \nonumber \\
C_2 & = &   - \frac{1}{(1-k)} \label{eq:constants}
\end{eqnarray}
The corresponding SM and MA surface participation values are dictated by the boundary conditions of the electric field,
and have the following form when all of the contamination layers possess identical relative dielectric constants,
$\epsilon_c$:
\begin{eqnarray}
r_{SM}\left(\frac{y}{a}\right) & \approx & \left(\frac{\epsilon_{sub}}{\epsilon_c}\right)^2 r_{SA}\left(\frac{y}{a}\right)
\label{eq:apprsm} \\
r_{MA}\left(\frac{y}{a}\right) & \approx & \left(\frac{1}{\epsilon_c}\right)^2 r_{SA}\left(\frac{y}{a}\right) \label{eq:apprma}
\end{eqnarray}
Participation within the entire volume of a contamination layer of thickness $\delta$ below the metallization layer can be approximated by integrating these equations with respect to $y$:
\begin{equation}\label{eq:apppi}
P_i \left(\frac{\delta}{a}\right) = \int_0^\delta r_i \left( \frac{y}{a} \right) dy
\end{equation}
to arrive at the expression in (\ref{eq:psa}) for $P_{SA}$.

Note that for a CPW structure, $K(k)$ must be replaced by $K'(k)$ in the denominator of the expressions for $\theta_a$ and $\theta_b$ (\ref{eq:thetaa}) and (\ref{eq:thetab}).  However, using $U_{tot}^{CPW}$ from (\ref{eq:utotcpw}) in the denominator of (\ref{eq:rsadef}) results in exactly the same representation of $r_{SA}, r_{SM}$ and $r_{MA}$ as shown in (\ref{eq:apprsa}), (\ref{eq:apprsm}) and (\ref{eq:apprma}), respectively.

\section{Estimation of dielectric loss tangents}\label{section:applosstan}

\begin{figure}[htbp!]
		 \centering
	\includegraphics[width=0.45\textwidth]{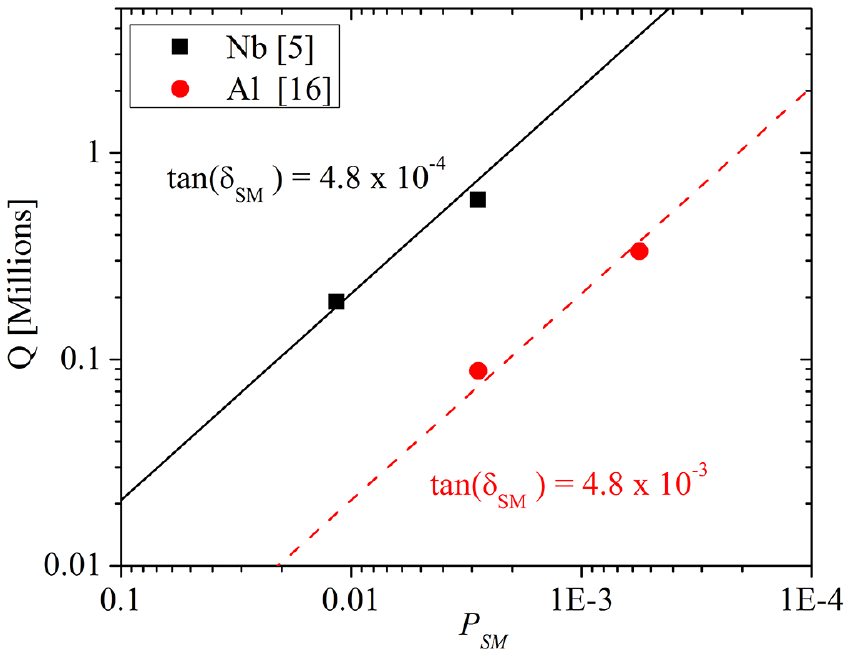}
		 \caption{Comparison of experimentally determined Q values with calculated surface participation values $P_{SM}$ for Nb-based qubits with Al-based junctions \cite{Gambetta16} and Al-based qubits \cite{Chang13}. The solid line corresponds to a loss tangent of tan($\delta_{SM}$) = $4.8 \cdot 10^{-4}$ and the dashed line to tan($\delta_{SM}$) = $4.8 \cdot 10^{-3}$ where it is assumed that SM surface participation is the dominant surface loss mechanism. }
		 \label{fig:tandcalc}
\end{figure}

The determination of loss tangent values for different surfaces was performed by comparing experimental Q values from qubits whose capacitor paddles were composed of sputter deposited Nb metallization with lift-off Al junction leads \cite{Gambetta16} or of completely lift-off Al metallization \cite{Chang13}.  SM participation values were calculated using the analytical approximation of (\ref{eq:psmapp}), where the paddles possessed an interdigitated comb structure with equal linewidths and gaps of 1 or 5 $\mu$m for the Nb-based qubits and 5 or 30 $\mu$m for the Al qubits.  As shown in Fig. \ref{fig:tandcalc}, if we assume that SM participation is the dominant mode of surface loss, then a loss tangent of $4.8 \cdot 10^{-4}$ produces Q values similar to that of the Nb qubits whereas a loss tangent of $4.8 \cdot 10^{-3}$ more closely resembles Q values associated with the Al qubits.

\bibliography{bibmaster}

\end{document}